\documentclass[twocolumn,showpacs,preprintnumbers,amsmath,amssymb]{revtex4}

\usepackage{graphicx}

\newcommand{\srf}{\ensuremath{S_{\mathrm{rf}}}}

\newcommand{\NEM}{\ensuremath{\mathrm{NEM}}}

\newcommand{\grscale}{0.5}

\newcommand{\unitvector}{\widehat}

\newcommand{\myvector}{\boldsymbol}
\newcommand{\rf}{rf}
\newcommand{\rw}{rw}

\begin{document}

\title{Theory of double resonance magnetometers based on atomic alignment}

\author{Antoine Weis}
\email{antoine.weis@unifr.ch}
\affiliation{%
Physics Department, University of Fribourg, Chemin du Mus\'ee 3,
1700 Fribourg, Switzerland
}%
\author{Anatoly S. Pazgalev}
\affiliation{Physics Department, University of Fribourg, Chemin du
Mus\'ee 3, 1700 Fribourg, Switzerland} \affiliation{Ioffe Physical
Technical Institute, Russ. Acad. Sc., St. Petersburg, 194021,
Russia}

\author{Georg Bison}
\affiliation{%
Physics Department, University of Fribourg, Chemin du Mus\'ee 3,
1700 Fribourg, Switzerland
}%

\date{May 26, 2006}

\begin{abstract}
We present a theoretical study of the spectra produced by
optical/radio-frequency double resonance devices, in which
resonant \emph{linearly} polarized light is used in the optical
pumping and detection processes. We extend previous work by
presenting algebraic results which are valid for atomic states
with arbitrary angular momenta, arbitrary \rf\ intensities, and
arbitrary geometries. The only restriction made is the assumption
of low light intensity. The results are discussed in view of their
use in optical magnetometers.
\end{abstract}

\pacs{32.60.+i, 32.30.Dx, 07.55.Ge, 33.40.+f}

 \maketitle

\section{\label{sec:intro}Introduction}
Since the 1950s the  combination of resonant optical excitation
and magnetic resonance has been an extremely valuable tool for
atomic spectroscopy. This double resonance technique
\cite{Brossel1952} has not only proven useful for investigating
atomic structure, for measuring properties of atoms, their
constituents, and their interactions, but has also led to
important applications in atom cooling, optically pumped frequency
standards, and optical magnetometers.

Magnetometers based on double resonance in atomic samples measure
the modulus of an externally applied magnetic field
$\myvector{B}_0$ via the Larmor precession frequency of the
sample's magnetization in that field \cite{blo62, Alex1992}. The
sample is typically a vapor of paramagnetic atoms (or diamagnetic
atoms excited to a metastable state with an orbital angular
momentum) sealed in a glass cell. A macroscopic magnetization is
created in the vapor by optical pumping with polarized resonance
radiation. The magnetization precesses in the magnetic field
$\myvector{B}_0$ to be measured (referred to as the offset field)
and that precession is driven by a (much weaker) magnetic field
$\myvector{B}_1(t)$ (referred to as the \rf\ field), co-rotating
with the magnetization around the offset field.

Since the optical properties of the medium, characterized by its
complex index of refraction, depend on its spin polarization, the
driven magnetization will induce periodic modulations of those
properties \cite{dehmeltmod, Happer1970}, which are then detected.
In most applications, the same light beam used to polarize the
medium is also used to monitor the oscillations by measuring
either the power or the polarization state of the transmitted
beam. The frequency of the induced oscillations coincides with the
oscillation frequency $\omega$ of the \rf\ field, and their
amplitude depends in a resonant way on the detuning between
$\omega$ and the Larmor frequency $\omega_0=\gamma_F B_0$
associated with the offset field. The Land\'e factor $\gamma_F =
g_F \mu_B / \hbar$ is characteristic for the pumped atomic state
with total angular momentum $F$.

Most practical double resonance devices rely on atomic orientation
prepared by optical pumping with \emph{circularly} polarized
light. In this paper we present a theoretical study of the
resonance signals obtained in double resonance spectroscopy using
\emph{linearly} polarized light. As shown first by Bell and Bloom
\cite{BellBloom61} magnetic resonance in aligned media leads to
signal modulations at the fundamental and at the second harmonic
of the \rf\  frequency. We derive algebraic expressions for the
spectral line shapes of the in-phase and quadrature components of
both signals and their orientation dependence. Previous
theoretical treatments of such signals
\cite{Gilles1992,Gilles1997} were restricted to specific angular
momentum states ($J=1$) or to low \rf\  powers. The results
presented here are more general in the sense that they apply to
arbitrary spin systems and that they are valid for arbitrary \rf\
power levels and for arbitrary orientations of $\myvector{B}_0$
with respect to the light polarization.

\section{Polarized atomic media}

\subsection{\label{sec:nultipole}Atomic Multipole Moments}

The density matrix $\rho $ of an ensemble of polarized atoms with
angular momentum $F$ can be expressed in terms of atomic multipole
moments $m_{k,q}$ according to \cite{blum}
\begin{equation}
\rho =\sum_{k=0}^{2F}\sum_{q=-k}^{k}m_{k,q}T_{q}^{(k)}\,,
\label{eq:tensdecrho}
\end{equation}
where the $T_{q}^{(k)}$ are standard irreducible tensor operators
\begin{eqnarray}
  \displaystyle
  T_{q}^{(k)}(F)&=&\displaystyle\sum_{M=-F}^{F} \displaystyle\sum_{M^{\prime} =-F}^{F}(-1)^{F-M^{\prime }}
  \\\nonumber
  && \times \left\langle F,M,F,-M^{\prime }|k,q\right\rangle  \,\left| F,M\right\rangle \left\langle F,M^{\prime }\right|
\end{eqnarray}
constructed from the angular momentum states $\left|
F,M\right\rangle $, and where the multipole moments $m_{kq}$ are
defined by
\begin{equation}
m_{k,q}=\left\langle T_{q}^{(k)\dag }\right\rangle =Tr(\rho
T_{q}^{(k)\dag })\,.
\end{equation}
The three multipole moments $m_{k=1,q=-1,0,+1}$ represent the
orientation of the medium, while the five components
$m_{k=2,q=-2,\ldots,+2}$ represent its alignment. The multipole
moments $ m_{k,q=0}$ are called longitudinal multipole moments and
their value depends only on sublevel populations. The multipole
moments $m_{k,q\neq 0}$ represent sublevel coherences and are
called transverse moments. The representation of the atomic
polarization in terms of multipole moments has a significant
advantage over a representation in terms of sublevel populations
and coherences. In principle, both representations require the
same number of parameters for the complete description of the
atomic ensemble. However, because electric dipole radiation
couples only \cite{Happer72}  to orientation ($k=1$) and alignment
($k=2$) it is sufficient to specify the corresponding $3+5=8$
multipole moments for the complete description of the system's
optical properties. Moreover, specific light polarizations couple
only to specific subsets of these 8 multipole moments, so that the
use of the tensor formalism in systems with large angular momenta
leads to a significant simplification of the mathematical
treatment. In the case discussed here only the (real) multipole
moment $m_{2,0}$ will be relevant. This approach therefore allows
one to derive results valid for systems with arbitrary angular
momenta.

A resonant \emph{circularly} polarized laser beam interacting with
an unpolarized atomic sample will create orientation ($k=1$,
vector polarization) and alignment ($k=2$, tensor polarization) in
the sample by optical pumping. The lowest order multipole that a
\emph{linearly} polarized light field can create is an atomic
alignment. While only atomic states with $J \geq 1/2$ can be
oriented, the condition $J\geq 1$ has to be fulfilled for the
creation of an aligned state. Note that an alignment along the
direction of light propagation can also be produced  by pumping
with \emph{unpolarized} light \cite{Colegrove1960,BellBloom61}.
The ground state of alkali atoms has an electronic angular
momentum $J=1/2$, which cannot be aligned. However, the hyperfine
interaction with the nuclear spin splits the ground state into two
hyperfine levels with total angular momenta $F_\pm=I\pm J$, which
can be aligned provided $F \geq 1$. An alignment can therefore be
prepared and/or detected only if the light source has a sufficient
spectral resolution to excite a single hyperfine transition. In
general the Doppler (and pressure) broadened spectra of discharge
lamps, used in conventional optically pumped magnetometers (OPM),
cannot be used to address individual hyperfine lines and hence do
not allow one to create nor to detect a ground state alignment.
However, radiation from a narrowband laser can resolve the
hyperfine structure and it is well known that a linearly polarized
laser beam can create an atomic alignment.

\subsection{\label{sec:drom dram}DROMs and DRAMs}
We will refer to an OPM based on atomic orientation as DROM
(double resonance orientation magnetometer), while we will speak
of DRAM (double resonance alignment magnetometer) when the
magnetization has the symmetry of an atomic alignment. Most of
the past research work on double resonance spectroscopy dealt with
oriented vapors, although alignment induced by (unpolarized) lamp
pumping in $J=1$ metastable states of $^4$He was already reported
in 1960/61 \cite{Colegrove1960,BellBloom61}. In alkali atoms,
alignment produced by lamp pumping can be observed using line
splitting by the quadratic Zeeman effect \cite{AlexAlign1990} or
isotope filtering \cite{AlexAlign1990,Happer1970}. The latter
technique is, however, restricted to Rb and cannot be applied to
other alkalis. As mentioned above, linearly polarized laser
radiation is an efficient means for producing alignment and a
discussion of linearly polarized laser pumping in metastable
$^4\text{He}$ can be found in
\cite{Gilles1992,Gilles_He4LsMag_2001}. These authors have
investigated several magnetometry techniques using both
orientation and alignment signals and they observed magnetic
resonances involving alignment using \rf\ fields, light intensity
modulation, polarization modulation, and frequency modulation.
A variant of the latter technique -- in which the transmitted
lights' polarization, instead of intensity, was measured -- was
realized with $^{87}$Rb \cite{Budker2000,BudkerPRA2002}.

\section{\label{sec:modelcalc}Model calculations}

\subsection{\label{sec:geom}DRAM geometry}

\begin{figure}
\includegraphics[width=8.5 cm]{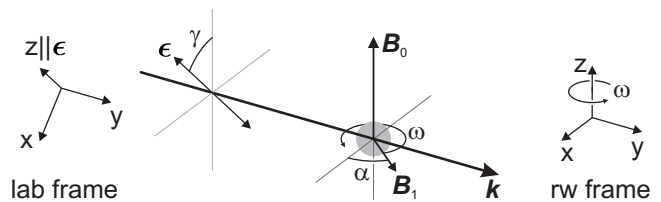}
\caption{\label{fig:AlphaGamma} Parametrization of the DRAM
geometry, in which  $\myvector{B}_1$ is perpendicular to
$\myvector{B}_0$. The rotating wave (rw) frame and the rotating
magnetic field $\myvector{B}_1$ are shown for one particular
moment in time $(t=0)$. }
\end{figure}

We will restrict the discussion to geometries in which the \rf\
field $\myvector{B}_{1}$ is perpendicular to the offset field
$\myvector{B}_{0}$. Because the signals are independent of the
orientation $\unitvector{k}$, the geometry of the problem is fully
determined by $\unitvector{B}_{0}$,\,$\unitvector{B}_{1}$, and
$\unitvector{\epsilon}$ so that one can consider the
parametrization shown in Fig.~\ref{fig:AlphaGamma}, in which
$\gamma$ denotes the angle between the offset field and the light
polarization \cite{Gilles1997}. The orientation dependence of the
signal amplitudes is given by $\gamma$, while the polar angle
$\alpha$ will lead to a phase shift in the time dependence of the
oscillating signals (Sec. \ref{sec:phase}).

In Fig.~\ref{fig:AlphaGamma} we have applied the rotating wave
approximation by decomposing the \rf\  field into two
counter-rotating fields, from which we have retained only the
component co-rotating with the alignment. This component is shown
in its position at time $t=0$, thereby defining the time origin of
the phases of the oscillating signal components. The magnetometer
signals are calculated following the 3-step (prepare, evolve,
probe) approach introduced in \cite{Weis1993, Kanorsky1993} and
discussed in detail in \cite{BudkerWeisRMP}. In the first step of
this model (preparation) one assumes the existence of a given
alignment in the atomic medium, without specifying how this
alignment was created. Details of the preparation process (optical
pumping, collisions with electrons or ions, spin exchange, etc)
thus do not need to be known. In the second step (magnetic
resonance) the initial alignment is allowed to evolve towards a
steady state value determined by the interaction with the external
fields and relaxation. Finally one considers in the third step
(probing) how the steady state alignment affects a linearly
polarized light beam traversing the medium. Strictly speaking this
approach is only valid for pump-probe experiments, in which the
atoms interact with spatially or temporally separated light fields
and where the equilibrium of step 2 is reached ''in the dark``.
However, as shown previously \cite{Weis1993, Kanorsky1993} for
level crossing signals the results obtained from the 3-step
approach give an excellent description of experimental findings if
the light intensity is sufficiently weak. Limitations of the model
will be addressed in section \ref{sec:validity}.

\subsection{\label{sec:step1}Step 1: Alignment creation}

We describe the alignment created by the preparation process in a
coordinate frame where the quantization axis lies along the light
polarization (lab frame in Fig.~\ref{fig:AlphaGamma}). In that
frame the only non-vanishing alignment component created by
optical pumping with \emph{linearly }polarized light is the
longitudinal multipole moment $m_{2,0}^\mathrm{ini}$ which can be
expressed in terms of the sublevel populations $p_M$ as
\begin{equation*}
m_{2,0}^\mathrm{ini}=N_2(F)\sum_{M=-F}^{F}\,p_{M}\left[3M^{2}-F(F+1)\right]
\,,
\end{equation*}
where $N_2(F)$ is a normalization constant \cite{blum}. In the
presence of an offset field $\myvector{B}_0$ the alignment
components perpendicular to $\myvector{B}_0$ will relax to zero
yielding a steady state value of
\begin{equation}
m_{2,0}^\mathrm{eq}=m_{2,0}^\mathrm{ini}\,\frac{3
\cos^2\gamma-1}{2}
\end{equation}
for the alignment along the magnetic field, given by the
projection of $m_{2,0}^\mathrm{ini}$ on the field direction. Note
that this steady state is reached only when the  Larmor frequency
is much larger than the transverse relaxation  rates. This
condition is well fulfilled for high-$Q$ magnetic  resonances
while it is not met by zero-field level-crossing  resonances
(ground state Hanle effect, nonlinear Faraday effect
\cite{BudkerWeisRMP}).

\subsection{\label{sec:step2}Step 2: Magnetic resonance}

This step describes the magnetic resonance process, i.e., the
evolution of the alignment under the combined actions of the
magnetic fields $\myvector{B}_0$, $\myvector{B}_1$, and relaxation
processes. It is described in a coordinate frame, which is related
to the lab frame by a static rotation of $-\gamma$ around the
$y$-axis and then by a dynamic rotation, at the frequency $\omega
$, around the new $z$-axis. In this frame, which we call the
rotating wave frame (\rw\ frame), the offset field
$\myvector{B}_0$ is along $z$, while the \rf\ field
$\myvector{B}_1$ appears to be static and oriented at an angle
$\alpha$ with respect to the $x$-direction (see Fig
\ref{fig:AlphaGamma}). This is the usual field configuration for
describing magnetic resonance processes. Note that
$m_{2,0}^\mathrm{eq}$ is not affected by the transformation to the
\rw\ frame. Due to the rotation of the coordinate frame, a
fictitious magnetic field $\myvector{B}_f=-\omega
\unitvector{\text{z}}/\gamma _{F}$ appears in the \rw\ frame, and
the atoms see a total field
$\myvector{B}_{\mathrm{tot}}=B_{1}\cos\alpha\,\unitvector{\text{x}}+B_{1}\sin\alpha\,\unitvector{\text{y}}+(B_{0}-\omega
/\gamma )\,\unitvector{\text{z}}$.

The evolution of the the system's density matrix is described by
the Liouville equation

\begin{equation}
\frac{d}{dt}{\rho }=\frac{1}{\hbar\, i}\left[ H,\rho \right] -\rho
_{\mathrm{relax}}\,, \label{eq:Liouville}
\end{equation}
with $H=-\myvector{\mu}\cdot \myvector{B}_{\mathrm{tot}}$, and
where $\rho_\mathrm{relax}$ describes the relaxation processes.
Inserting the multipole decomposition (\ref{eq:tensdecrho}) into
(\ref{eq:Liouville}) yields the following equations of motion for
the multipole moments $m_{2,q}$
\begin{equation}
\frac{d}{dt}m_{2,q}=\sum_{q^{\prime }}\mathbb{O}^{(2)}_{qq^{\prime
}}m_{2,q^{\prime }}-m_{2,q}^\mathrm{relax} \quad q= -2,-1,\dots, 2
\label{eq:AlignBloch}\,,
\end{equation}
where $\mathbb{O}^{(2)}_{qq^{\prime }}$ is given by
\begin{equation*}
\mathbb{O}^{(2)}_{qq^{\prime }} = \left(
{%
\setlength{\arraycolsep}{0 mm}
\begin{array}{ccccc}
  -2i\delta  & i\omega _{1} p_-& 0 & 0 & 0 \\
  i\omega _{1} p_+ & -i\delta  & i\sqrt{\frac{3}{2}}\omega _{1}p_- & 0 & 0 \\
  0 & i\sqrt{\frac{3}{2}}\omega _{1} p_+& 0 & i\sqrt{\frac{3}{2}}\omega_{1}  p_-&  0 \\
  0 & 0 & i\sqrt{\frac{3}{2}}\omega _{1} p_+ & i \delta  & i\omega_{1} p_- \\
  0 & 0 & 0 & i\omega _{1} p_+& 2i \delta
\end{array}
} %
\right) ,
\end{equation*}
in which $\omega _{1}=\gamma_{F}B_{1}$ is the Rabi frequency of
the \rf\ field, $p_\pm=\exp(\pm i \alpha)$ are phase factors that
describe the orientation of $\omega _{1}$ in the $xy$-plane, and
$\delta=\omega-\omega_{0}$ is the detuning of the radio frequency
$\omega $ with respect to the Larmor frequency $\omega _{0}$. The
relaxation terms are given by
\begin{equation}
\left(
\begin{array}{c}
m_{2,2}^\mathrm{relax} \\
m_{2,1}^\mathrm{relax} \\
m_{2,0}^\mathrm{relax} \\
m_{2,-1}^\mathrm{relax} \\
m_{2,-2}^\mathrm{relax}
\end{array}
\right) =\left(
\begin{array}{l}
\Gamma _{2}\,m_{2,2} \\
\Gamma _{1}\,m_{2,1} \\
\Gamma _{0}\,(m_{2,0}-m_{2,0}^\mathrm{eq}) \\
\Gamma _{1}\,m_{2,-1} \\
\Gamma _{2}\,m_{2,-2}
\end{array}
\right) \,.  \label{eq:AlignRelax}
\end{equation}
where we assume that the multipole moments $m_{2,q}$ relax at
rates $\Gamma_{|q|}$. The transverse alignment components
$m_{2,q\neq0}$ relax towards a zero value, while the longitudinal
component $m_{2,0}$ relaxes towards $m_{2,0}^\mathrm{eq}$,
introduced in step 1. Below, we present very general results for
the case when all three relaxation rates are different, although
it is well known that there are specific relations between those
rates when the relaxation mechanism has specific rotational
symmetries \cite{HapperRelax}. Eqations \ref{eq:AlignBloch} are a
generalization of the well known Bloch equations
\begin{equation}
\frac{d}{dt}m_{1q}=\sum_{q^{\prime }}\mathbb{O}^{(1)}_{qq^{\prime
}}m_{1q^{\prime }}-m_{1,q}^\mathrm{relax}\quad q= -1, 0, 1 \,,
\label{eq:OrientBloch}
\end{equation}
describing the evolution of the three orientation components
$m_{1,q}$.

We have used a computer algebra software to determine algebraic
expressions for the steady state (${d}/{dt}\,\, m_{2,q}=0$)
solutions $m_{2,q}$ of (\ref{eq:AlignBloch}). These solutions are
then transformed back to the lab frame, by first applying a
dynamic rotation at the rate $-\omega $ around the $z$-axis, and
then a static rotation by $\gamma $ around the $y$-axis of the
\rw\ frame. In this way one can derive algebraic expressions for
the time dependent multipole moments $ m_{2,q}(t)$ in the lab
frame.

\subsection{\label{sec:step3}Step 3: Alignment detection}

In the third and final step, we calculate the effect the time
dependent multipole moments have on the optical absorption
coefficient of the medium. One can show that the absorption
coefficient of a medium described by $m_{k,q}$ for linearly
polarized light is proportional to
\begin{equation}
\kappa \propto \frac{A_{0}}{\sqrt{3}}m_{0,0}-\sqrt{\frac{2}{3}}%
A_{2}m_{2,0}\,,  \label{eq:kappa1}
\end{equation}
where the (analyzing power) $A_{k}$ depends only on the states
$\left| n_{g},L_{g},J_{g},F_{g}\right\rangle $ and $\left|
n_{e},L_{e},J_{e},F_{e}\right\rangle $ coupled by the light. The
multipole moments in (\ref{eq:kappa1}) are defined with respect to
a quantization axis oriented along the incident light
polarization, which is the case in the lab frame, i.e., the frame
in which the results of step 2 are expressed. The monopole moment
$m_{0,0}$ describes the total population of the hyperfine ground
state $\left| n_{g},L_{g},J_{g},F_{g}\right\rangle $. We assume
the optical transition to be closed and the light intensity to be
so weak that excited state populations remain negligible, so that
the monopole moment does not depend on time. The only time
dependent (oscillating) component of the absorption coefficient is
therefore proportional to $m_{2,0}(t)$. We define the time
dependent DRAM signal, normalized to the longitudinal alignment
initially produced by the optical pumping, as
\begin{equation}
\label{eq:DRAMsignal} S(t)=\frac{m_{2,0}(t) } {
m_{2,0}^\mathrm{ini} }\,.
\end{equation}

\subsection{\label{sec:validity}Validity of the three step approach}

The three step approach is  only valid if steady state conditions
are reached in steps 1 and 2. This is fulfilled when the pump rate
$\Gamma_p$, at which alignment components are modified by the
interaction with the light is negligible compared to the
relaxation rates $\Gamma_{|q|}$. This condition can be realized
experimentally at low light powers, however, at the cost of a
decreased signal to noise ratio. OPMs are known to perform best
when $\Gamma_p$ is comparable to $\Gamma_{|q|}$. The lowest order
correction, taking the depolarization of light interactions into
account, can be described by
\begin{equation}
\label{eq:gammap} \Gamma_{|q|}\rightarrow \Gamma_{|q|}+\Gamma_p =
\Gamma_{|q|}+ \eta P_L\,,
\end{equation}
where $P_L$ is the laser power.

It is well known that substitution (\ref{eq:gammap}) is valid to
all orders in $P_L$ for a DROM in a spin 1/2 system \cite{ogawa},
in which orientation is the only multipole moment that can be
created. It is reasonable to assume that the same statement can be
made for a DRAM in a spin 1 system in which alignment is the only
multipole moment created and detected by the linearly polarized
light. For angular momenta $F>1$ the creation of higher order
($k>2$) multipole moments and their transfer back to (detectable)
$k=2$ moments limits the validity of substitution
(\ref{eq:gammap}) to low light powers.

\section{\label{sec:results}Results}

\begin{figure*}
\includegraphics[scale=\grscale]{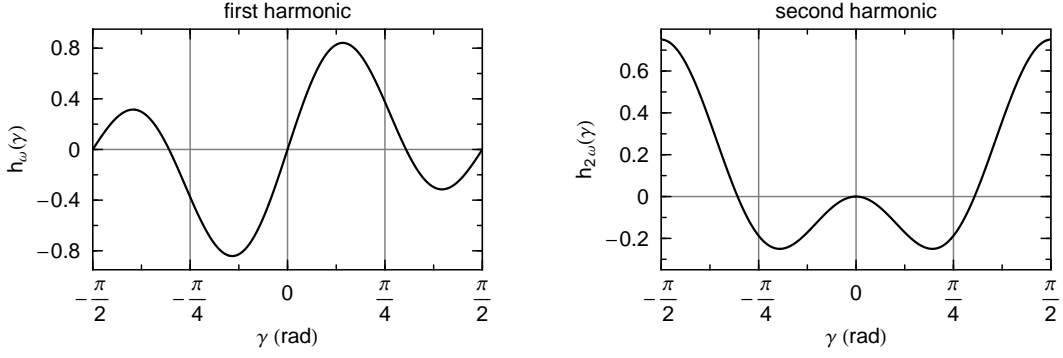}
\caption{Angular dependence of the first ($h_{\protect\omega }$,
left) and second ($h_{2\protect\omega }$, right) harmonic signals on the angle $\protect%
\gamma $ between the light polarization $\myvector{\epsilon}$ and
the magnetic offset field $\myvector{B}_0$. } \label{fig:hom2om}
\end{figure*}

The calculation outlined above yields signals $S_\omega(t)$ and
$S_{2\omega}(t)$ which are modulated at the \rf\ frequency
$\omega$ and at its second harmonic $2\omega$, and which can be
written as
\begin{subequations}
\label{eq:signals}
\begin{eqnarray}
  S_{\omega }(t)&\!\!\!=
  \phantom{_2} h_{\omega}(\gamma) \left[ \right.&
    \!\!\! \phantom{-}D_{\omega }\,\cos \left(\omega t-\alpha\right)  \\\nonumber
  && \!\!\! \left. -  A_{\omega }\,\sin \left(\omega t-\alpha\right)\right] \,,
\label{subeq:somega}\\
  S_{2\omega }(t)&\!\!\!=h_{2\omega }(\gamma )\left[\right. &
  \!\!\!           -A_{2\omega }\,\cos \left(2\omega   t-2\alpha\right) \\  \nonumber
  && \!\!\! \left. -D_{2\omega }\,\sin \left(2\omega  t-2\alpha\right)\right] \,, \quad \phantom{,}
  \label{subeq:s2omega}
\end{eqnarray}
\end{subequations}
where the angular dependence of the signals $h_{\omega }(\gamma )
$ and $h_{2\omega }(\gamma )$ (Fig.~\ref{fig:hom2om}) is given by
\begin{subequations}
\label{eq:angdep}
\begin{eqnarray}
 h_{\omega }(\gamma ) &=&\frac{3}{2}\sin \gamma \cos \gamma \left( 3\cos ^{2}\gamma -1\right) \label{subeq:hgamma1}
\\%
 h_{2\omega }(\gamma ) &=&\frac{3}{4}\sin ^{2}\gamma \left( 1-3\cos ^{2}\gamma \right)\,. \label{subeq:hgamma2}
\end{eqnarray}
\end{subequations}

As stated earlier the orientation angle $\alpha$ of the \rf\ field
appears as a phase shift. The in-phase and quadrature components
of the signal have both absorptive ($A_{\omega}$,  $A_{ 2\omega
}$) and dispersive ($ D_{\omega}$, $D_{2\omega }$) lineshapes
given by
\begin{subequations}
\label{eq:lineshapes3gamma}
\begin{eqnarray}
D_{\omega }&\!\!=&\!\!\frac{
\Gamma _0 \omega_1 \left(\Gamma_2^2+4 \delta^2-2 \omega_1^2
\right) \delta
}{Z} \, ,
 \\
A_{\omega } &\!\!=&\!\!\frac{
\Gamma _0 \omega_1 \left[\left(\Gamma_2^2 + 4 \delta^2\right)
\Gamma_1 + \Gamma_2 \omega_1^2 \right]
}{Z}\, ,
\\
\label{subeq:D2omega3gamma}
 D_{2\omega } &\!\!=&\!\!\frac{
\Gamma_0 \omega_1^2 \left(2 \Gamma_1+\Gamma_2 \right) \delta
 }{Z} \,,   \\
\label{subeq:A2omega3gamma}
A_{2\omega } &\!\!= %
&\!\!\frac{
\Gamma _0 \omega_1^2 \left(\Gamma_1\Gamma_2 - 2 \delta^2 +
\omega_1^2\right)
}{Z} \, ,
\end{eqnarray}
\end{subequations}
with
\begin{eqnarray}
Z&=& \Gamma_0 \left(\Gamma_1^2+\delta^2\right) \left(\Gamma _2^2+
4 \delta ^2\right)
\nonumber \\ && %
  +\left(\Gamma _1 \Gamma _2 \left(2 \Gamma _0+3
   \Gamma _2\right)
   - 4 \delta ^2 \left(\Gamma _0-3 \Gamma _1\right)\right) \omega _1^2
\nonumber \\ && %
   + \left(\Gamma _0+3 \Gamma _2\right) \omega _1^4
\, ,\label{eq:denom3gamma}
\end{eqnarray}

Equations (\ref{eq:lineshapes3gamma}) and (\ref{eq:denom3gamma})
can be simplified substantially if we assume an isotropic
relaxation by setting $\Gamma_0=\Gamma_1=\Gamma _2\equiv \Gamma $.
We will stick to this assumption in the following discussion since
it does not change the general properties of the spectra. We
further simplify the notation by introducing a dimensionless \rf\
saturation parameter $\srf =( \omega_{1} / \Gamma ) ^{2}$ and a
normalized detuning $x=\delta/\Gamma$. With these assumptions and
definitions we obtain
\begin{subequations}
\label{eq:shapes}
\begin{eqnarray}
D_{\omega }(x,\srf )&\!\!=&\!\!\frac{x\left( 1-2\srf
+4x^{2}\right) \,\sqrt{\srf } }{\left( 1+\srf +x^{2}\right) \left(
1+4\srf +4x^{2}\right) }\,,\qquad\phantom{,}
\label{subeq:Domega} \\
A_{\omega }(x,\srf ) &\!\!=&\!\!\frac{\left(1+\srf +4x^{2}\right)
\,\sqrt{\srf }}{\left( 1+\srf +x^{2}\right) \left( 1+4\srf
+4x^{2}\right) }\, ,
\label{subeq:Aomega}\\
 D_{2\omega }(x,\srf ) &\!\!=&\!\!\frac{3\,x \,\srf}{\left( 1+\srf +x^{2}\right) \left(
1+4\srf +4x^{2}\right) } \,,  \label{subeq:D2omega} \\
A_{2\omega }(x,\srf ) &\!\!=&\!\!\frac{\left(1+\srf
-2x^{2}\right)\,\srf}{\left( 1+\srf +x^{2}\right) \left( 1+4\srf
+4x^{2}\right) } \, . \label{subeq:A2omega}
\end{eqnarray}
\end{subequations}
In the following we will discuss in detail the different
properties of the signals (\ref{eq:signals}) with lineshapes
(\ref{eq:shapes}).

\subsection{\label{sec:shapes}Line shapes}

\begin{figure*}
\center
\includegraphics[scale=\grscale]{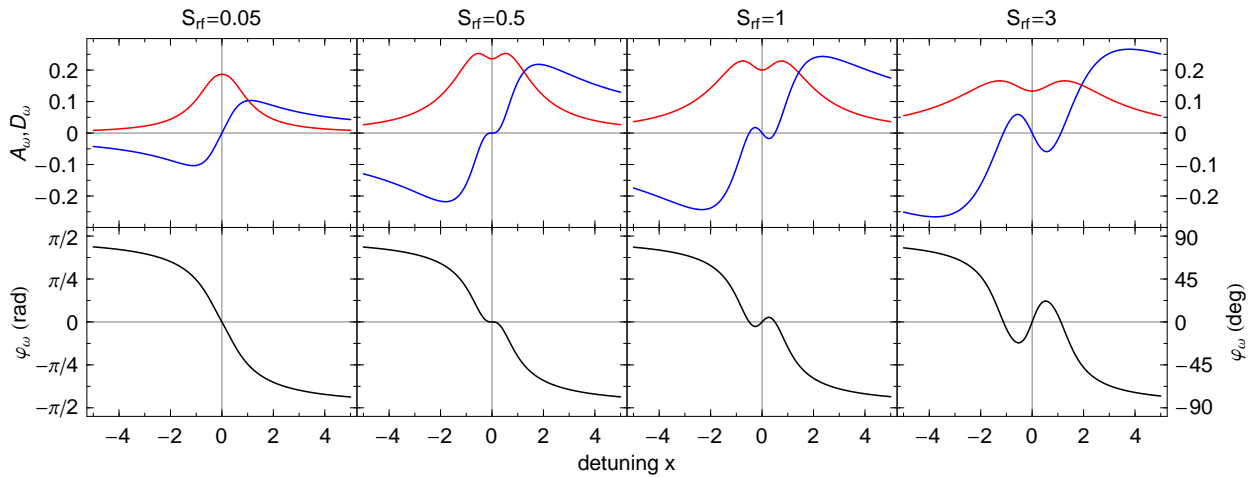}
\caption{(Color online) Top line: Line shapes of the absorptive
($A_{\protect\omega }$) and dispersive ($D_{\protect\omega }$)
components of the first harmonic signal for different values of
the \rf\  saturation parameter $\srf$.
Bottom line: Shape of the phase signal $\varphi_{\protect\omega}$
for the same values of $\srf$.
$x= (\omega-\omega_0)/\Gamma$ is the normalized detuning of the
\rf\ frequency $\omega$ with respect to the Larmor frequency
$\omega_0$. } \label{fig:shapesomega}
\end{figure*}

\begin{figure*}
\center
\includegraphics[scale=\grscale]{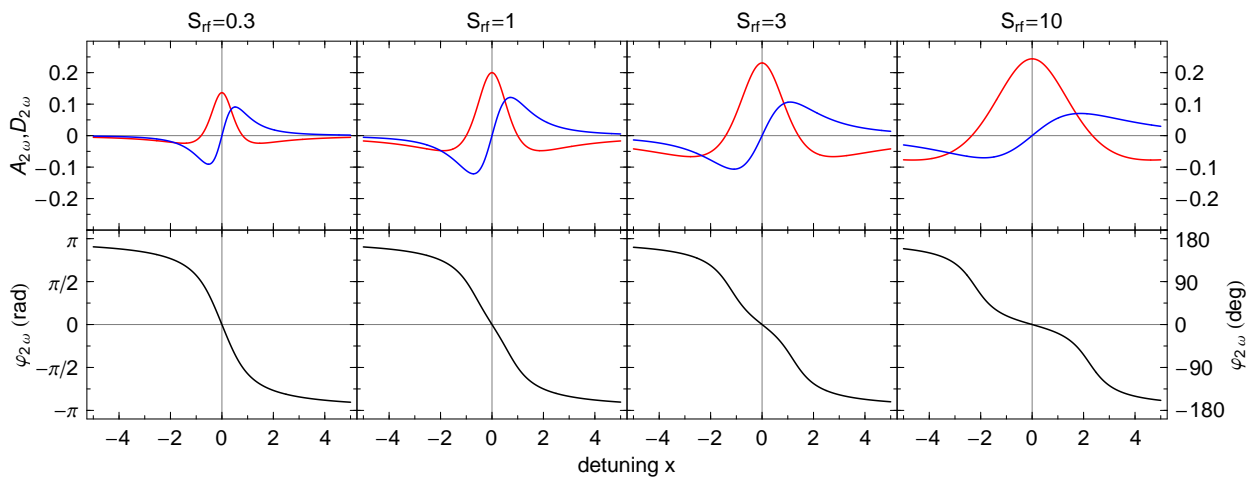}
\caption{(Color online) Top line: Line shapes of the absorptive
($A_{2\protect\omega }$) and dispersive ($D_{2\protect\omega }$)
components of the second harmonic signal for different values of
the \rf\  saturation parameter $\srf $.
Bottom line: Shape of the phase signal $\varphi_{2\protect\omega}$
for the same values of $\srf$. } \label{fig:shapes2omega}
\end{figure*}

For low \rf\  intensities, i.e., for $\srf \rightarrow 0$, the
lineshapes (\ref{eq:shapes}) reduce to
\begin{subequations}
\label{eq:omegalowrf}
\begin{eqnarray}
D_{\omega }(x,\srf \rightarrow 0)=\frac{x}{1+x^{2}}\sqrt{\srf }\,,
\label{subeq:DomegaLowrf$}
\\A_{\omega }(x,\srf \rightarrow 0)=\frac{1}{1+x^{2}}\sqrt{\srf } \label{subeq:AomegaLowrf$} \,,
\end{eqnarray}
\end{subequations}
and
\begin{subequations}
\label{eq:2omegalowrf}
\begin{eqnarray}
D_{2\omega }(x,\srf  &\rightarrow &0)=\frac{3\,x}{\left(
1+x^{2}\right)
\left( 1+4x^{2}\right) }\srf \, ,  \label{subeq:D2omegaLowrf} \\
A_{2\omega }(x,\srf  &\rightarrow &0)=\frac{1-2x^{2}}{\left(
1+x^{2}\right) \left( 1+4x^{2}\right) }\srf \, .
\label{subeq:A2omegaLowrf}
\end{eqnarray}
\end{subequations}
Expressions (\ref{eq:angdep})  and (\ref{eq:omegalowrf}),
correspond to results obtained in earlier work
\cite{Gilles1992,Gilles1997}.

The corresponding spectra can be seen in the leftmost columns of
Figs.~\ref{fig:shapesomega} and \ref{fig:shapes2omega}. These
figures also show how the lineshapes of the absorptive and
dispersive signals $A_{\omega },D_{\omega },A_{2\omega }$, and
$D_{2\omega }$ change with increasing \rf\ intensity. A narrow
additional spectral feature appears in the first harmonic signal
for $\srf >0.5$. The origin of this structure can be explained as
follows:\ the basic interaction of the \rf\  field with the
angular momentum is the coupling of adjacent Zeeman sublevels. The
corresponding $\Delta M=\pm 1$ coherences oscillate at the \rf\
frequency $\omega$ and their detection by the light field
constitutes the first harmonic signal. With increasing \rf\ power,
a further interaction of a $\Delta M=1$ coherence with the \rf\
field $B_{1}$ becomes possible and leads to the creation of a
$\Delta M=2$ coherence, whose oscillation produces the second
harmonic signal. An additional interaction of the $\Delta M=2$
coherence with the \rf\ field produces both $\Delta M=3$ and
$\Delta M=1$ coherences. While the former cannot be detected
optically since linearly polarized light couples at most to
$\Delta M=2$ coherences, the latter directly contributs to the
first harmonic signal. In this sense, the additional features can
be understood as resulting from the creation of a $\Delta M=2$
coherence by a second order interaction with the \rf\ field, the
evolution of that coherence in the offset field, and back-transfer
to a $\Delta M=1$ coherence by an additional interaction with the
\rf\  field followed by detection of that coherence at the first
harmonic frequency $\omega $.

The additional narrow feature in the absorptive signal that
appears at large \rf\  intensities was already observed in the
pioneering work by Colegrove and Franken on optically induced
alignment \cite{Colegrove1960}. Based on the results presented
above the feature can be explained in terms of line
superpositions. In fact, the expression for the absorptive first
harmonic signal $A_\omega$ (Eq.~\ref{subeq:Aomega}) can be
rewritten as
\begin{equation}
A_{\omega}=\frac{1+\srf }{1+\srf +x^{2}}\,\sqrt{\srf }-\,\frac{4
\srf }{1+4\srf +4x^{2}}\,\sqrt{\srf }\,, \label{eq:a2res}
\end{equation}
i.e., as a superposition of two absorptive Lorentzian line shapes
whose widths, for $\srf \rightarrow 0$, differ by a factor of 2,
while they become equal for large values of $\srf $. The
appearance of the central dip is a consequence of the different
rates at which the contributions saturate with increasing $\srf $.
At low \rf\ powers, the amplitudes of the two contributions to the
first harmonic signals (\ref{eq:a2res}) grow as $\srf^{1/2}
\propto \omega_1$ and $\srf^{3/2}\propto \omega_1^3$ respectively,
which reflects that these resonances correspond to first and third
order processes as discussed above. The second harmonic signals
(Eqs.~\ref{subeq:D2omegaLowrf}, \ref{subeq:A2omegaLowrf}), on the
other hand, grow as $\omega_1^2$, which reflects their second
order nature.

The dependence of the line widths, i.e., the frequency separation
$\Delta\omega_{\mathrm{FW}}$ of the maxima and minima of the
dispersive signals $D_\omega$ and $D_{2\omega}$ on $\srf $ can be
inferred from the derivatives of (\ref{subeq:Domega}) and
(\ref{subeq:D2omega}). The dispersive line shape of the second
harmonic signal $D_{2\omega}$ is thus 2.6 times narrower than the
corresponding first harmonic signal.

\subsection{The phase of the signals \label{sec:phase}}

We define the phases $\varphi_{\omega}$  ($ \varphi_{2\omega}$) of
the first and second harmonic signals as the phase difference
between $S_{\omega}(t)$ ($S_{2\omega}(t)$) and oscillations that
are proportional to $\cos \omega t$ ($\cos 2\omega t$)
respectively. This definition is motivated by the fact that the
\rf\ field in the lab frame is proportional to $\cos  \omega t$.
As an alternative to the parametrization of the signals in terms
of in-phase and quadrature components (Eq.\,\ref{eq:signals}) one
can write $S_{\omega}(t)$ and $S_{2\omega}(t)$ in terms of moduli
$R_\omega$, $R_{2\omega}$ and phases $\varphi_\omega$,
$\varphi_{2\omega}$ according to
\begin{subequations}
\label{eq:signalsRphi}
\begin{eqnarray}
  S_{\omega} &=& h_{\omega}(\gamma) R_{\omega}(x,\srf)
  \cos \left[\omega t  + \varphi_\omega(x,\srf)\right] \,, \\
    S_{2\omega} &=& h_{2\omega}(\gamma) R_{2\omega}(x,\srf)
  \cos \left[2 \omega t  + \varphi_{2
  \omega}(x,\srf)\right]\,,\qquad\phantom{.}
\end{eqnarray}
\end{subequations} with
\begin{eqnarray}
  \varphi_{\omega} & = &  \arctan\left(
    \frac{  A_{\omega }(x,\srf ) }{  D_{\omega }(x,\srf ) }
    \right) - \alpha \\
  \varphi_{2\omega} & = &  \arctan\left(
    - \frac{D_{2 \omega }(x,\srf ) }{A_{2 \omega }(x,\srf ) }
    \right) - 2 \alpha .
\end{eqnarray}

A dual-phase lock-in amplifier can be used to extract the
$\varphi$'s from the signal. Such amplifiers usually allow one to
apply an offset phase $\varphi_{\mathrm{os}}$ to the signal which
then adds to $\alpha$. For stabilization purposes, it is practical
to choose that offset phase such that the total phase vanishes at
the center of the resonance
$\varphi_{\omega}(x=0)=\varphi_{2\omega}(x=0) = 0$.
This choice avoids phase discontinuities and provides a signal
$\varphi(x) \propto x$ (for $x\ll 1$) that is proportional to
magnetic field changes near the center of the resonance. In the
parametrization defined above this can be realized simultaneously
for both signals  when $\alpha+\varphi_{\mathrm{os}}= \pi/2$. In
that case the phase signals take the form
\begin{subequations}
\label{eq:DRAMp}
\begin{eqnarray}
\varphi_\omega &=& -\arctan\left(x \frac{1-2 \srf  + 4x^2}{1 +
\srf +4x^2 } \right)\, ,   \label{eq:DRAMp1} \\
\varphi_{2 \omega} &=& -\arctan\left(\frac{ 3 x}{1 + \srf -2x^2 }
\right). \label{eq:DRAMp2}
\end{eqnarray}
\end{subequations}
Examples of $\varphi_\omega (x)$ and $\varphi_{2 \omega}(x)$  for
different  values of \srf\ are shown in Figs.
\ref{fig:shapesomega} and \ref{fig:shapes2omega} respectively.

The phase of the first harmonic signal  $\varphi_\omega$ (Eq.
\ref{eq:DRAMp1}) for low \rf\  power $(\srf \ll 1/2)$ is identical
to the phase of an orientation based magnetic resonance signal.
Conversely to that DROM phase, which does not depend on \srf, the
DRAM phases $\varphi_{\omega}$ and $\varphi_{2\omega}$ depend on
\srf, and $\varphi_{\omega}$ even changes the sign of its slope at
$\srf = 1/2$.
Note that $\varphi_{2 \omega}$ makes a total phase swing of $2
\pi$ across the resonance, while $\varphi_\omega$ swings only by
$\pi$.

\section{Magnetometry considerations}

\begin{figure*}
\includegraphics[scale=\grscale]{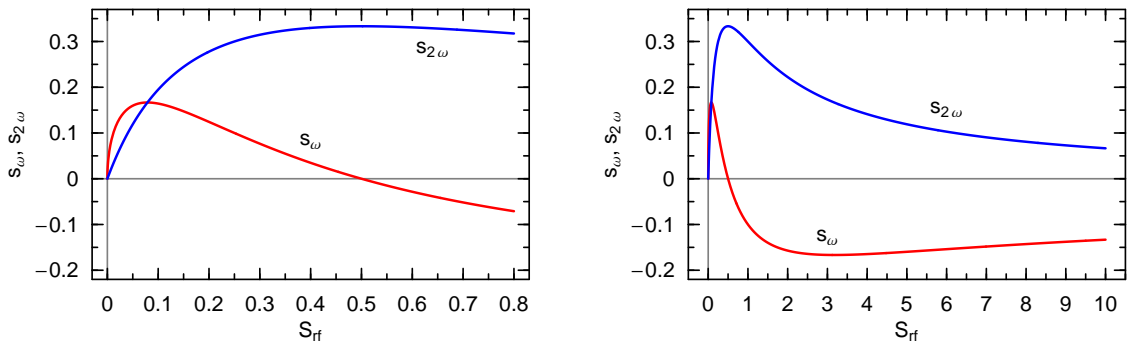}
\caption{(Color online) Slopes $s_{\omega, 2 \omega}$ of the
dispersive signals on resonance ($x=0$) as a function of the \rf\
saturation parameter for two different ranges of $\srf $.}
\label{fig:slopes}
\end{figure*}

There are two modes of OPM operation, the self-oscillating mode
and the free running mode, which differ in the way the magnetic
field value is extracted from the signals described above.
Self-oscillating (or phase stabilized) magnetometers use feedback
to keep the driving frequency $\omega$ equal to $\omega_0$, and in
such magnetometers, $\omega$ is measured with a frequency counter
\cite{GroegerSoundcard}.
We will not discuss the self-oscillating mode in detail since the
feedback is difficult to describe analytically.
Instead, we discuss the free-running mode which yields an
identical magnetic field sensitivity. The goal of the following
discussion is not the derivation of absolute field sensitivities,
but rather a comparison of the relative magnetometric
sensitivities that one can expect from the first and second
harmonic signals.

As a free-running magnetometer application we consider the
recording of signal changes induced by variations of the offset
field $B_0$ while the \rf\ frequency $\omega$ is kept constant.
Either the dispersive signals $D_{\omega}$ and $D_{2\omega}$
(Eqs.~\ref{subeq:Domega}\,,\ref{subeq:D2omega}) or the phase
signals $\varphi_{\omega}$ and $\varphi_{2\omega}$
(Eqs.~\ref{eq:DRAMp1}\,,\ref{eq:DRAMp2}) can be used as
discriminating signals as they both feature a linear dependence on
$B_0$ changes near the center of the resonance ($x=0$).

The resolution with which field changes can be detected is limited
by noise processes such as photon shotnoise, electron shotnoise
(in a photodiode), and spin projection noise, all of which have a
white noise spectrum.
We specify the magnetometric sensitivity in terms of the noise
equivalent magnetic field (\NEM) \cite{georgJOSA,FRAPLsOPM}, which
is the amplitude $\Delta B^{\NEM}$ of field fluctuations which
induce fluctuations of the (dispersive) signal $S(B_0)$ that are
equal to the signal noise $\Delta S$:
$$
\Delta B^{\NEM}=\Delta S\,\left( \left. dS/dB_0\right|
_{x=0}\right)^{-1}
$$
One can show that for a given noise level of the modulated signal
one obtains the same \NEM\ either from the demodulated signal
$S=D_\omega (D_{2\omega})$ or from the demodulated signal
$S=\varphi_\omega (\varphi_{2\omega})$. The minimal value of
$\Delta B^{\NEM}$ is thus obtained under conditions which maximize
$h_{\omega}\cdot s_{\omega}$ and $h_{2\omega} \cdot s_{2\omega}$,
where the (on resonance) slopes $s_{\omega}$ and $s_{2\omega}$ are
given by

\begin{subequations}
\label{eq:Dslopes}
\begin{eqnarray}
s_\omega = \phantom{_2} \left. \frac{\text{d}D_{\omega }}{\text{d}x}\right| _{x=0} &=&\frac{%
(1-2\srf )\sqrt{\srf }}{(1+\srf )(1+4\srf )}\,, \\
s_{2 \omega} = \left. \frac{\text{d}D_{2\omega }}{\text{d}x}\right| _{x=0} &=&\frac{3\srf %
}{(1+\srf )(1+4\srf )}\,.
\end{eqnarray}
\end{subequations}
Figure \ref{fig:slopes} shows their dependence on the \rf\
intensity. The zero crossing of the slope of the first harmonic
signal at $\srf =0.5$ marks the emergence of the narrow central
feature in Fig.~\ref{fig:shapes2omega}.

The maximal sensitivity (minimal $\Delta B^{\NEM}_{\omega, 2\omega
}$) is achieved by choosing a geometry which which maximizes
$h_{\omega}$ and $h_{2\omega}$ and an \rf\  intensity, which
maximizes $s_{\omega}$ and $s_{2\omega }$. For the first harmonic
signal one finds $\max [ h_{\omega } \cdot s_{\omega}] =0.141$ for
$\gamma=25.5$ degrees and $\srf=0.079$, while for the second
harmonic signal one has $\max [ h_{2 \omega } \cdot s_{2 \omega}]
=0.25 $ for $\gamma=90$ degrees and $\srf=0.5$. Under optimized
geometrical and \rf\ power conditions and for a given noise level
the second harmonic signal  is thus expected to yield a $1.8$
times higher sensitivity to magnetic field changes than the first
harmonic signal.

\section{Summary and Conclusion}
We have presented a general theoretical framework for the
calculation of optical \rf\ double resonance signals that can be
easily applied to oriented or aligned atomic media. The theory
yields analytical results with a broad range of validity, and is
only limited by the assumption of low light power.

DRAMs, i.e., magnetometers that use linearly polarized light,
present several potential advantages over the well known DROM
scheme:
\begin{itemize} %
\item Line widths: The line shapes of the second harmonic DRAM
signal have significantly narrower linewidths than the DROM signal
under identical conditions. Narrow linewidths potentially increase
the magnetometric sensitivity and suppress systematic errors in
optical magnetometers due to long term baseline drifts
\cite{Alex_KOPM4q}.
\item Light shift: In DROM devices the interaction of the atoms
with the circularly polarized laser light leads to $M$ dependent
energy shifts (vector light shift) of the sublevels
$\left|F,M\right>$ when the laser frequency is not centered on the
optical resonance line. In that case, power and frequency
fluctuations of the laser mimic magnetic field fluctuations,
thereby limiting the magnetometric performance. In the DRAM device
the linearly polarized light produces a tensor light shift
\cite{HapperLS68} depending on $M^2$, which does not have the
characteristics of the Zeeman interaction and will therefore not
affect the magnetometric performance.
\item Geometry: The DROM scheme achieves a maximal sensitivity for
$\theta=\pi/4$. The 45 degree angle that the laser beam has to
make with the magnetic field seriously limits applications which
call for a compact arrangement of multiple sensors. In
multi-channel devices, as required, e.g., for cardiomagnetic
measurements \cite{georg2}, the use of the DRAM signals offers the
advantage that the laser beam can be oriented either parallel or
perpendicular to the offset field.
\item Vector magnetometry: Both the DROM and the DRAM devices are
scalar magnetometers and their resonance frequency measures
$|B_0|$. However, the DRAM signals can be used to realize a vector
magnetometer, since the ratio of $R_{2\omega}/R_{\omega}$ is
proportional to $\tan\gamma$, for all values of \srf. The
knowledge of $|\myvector{B_0}|$ and $\gamma$ locates
$\unitvector{B}_0$ on a cone and the variation of the signal with
$\phi$, obtained by rotating the polarizer, will determine the
polar angle of $\widehat{B}_0$ on the cone. In this way the DRAM
scheme can be used to infer all three vector components of the
field.
\item Relaxation: We are in the process of performing extensive
experimental studies of the DRAM properties in paraffine coated
cesium cells \cite{DRAMExp}. First results indicate that a
description of the signals using three independent relaxation
rates (Eqs.~\ref{eq:lineshapes3gamma}) is required to describe the
experimental lineshapes in detail. The DRAM signals thus seem to
offer a convenient way for studying spin relaxation processes in
aligned media.
\item Signal noise: Diode lasers are convenient light sources for
double resonance experiments. However, they often feature a $1/f$
 (flicker) intensity noise at low frequencies that turns
into the white shot noise level at higher frequencies. The
detection of $D_{2\omega}$ and $A_{2\omega}$ at twice the Larmor
frequency makes it easier to operate in a region where the laser
noise is less affected by flicker noise.
\end{itemize} %

Cs OPMs in the DROM geometry have a shot noise limited sensitivity
of 10 fT/Hz$^{1/2}$ \cite{FRAPLsOPM}. A direct experimental
comparison of DRAM and DROM magnetometers is currently underway in
our laboratory. This study will show if the potential adventages
can be realized in practice.

\begin{acknowledgments}
The authors thank P.~Knowles and G.~Di~Domenico for useful
discussions and a critical reading of the manuscript. One of us
(A.~S.~P.) acknowledges financial support by the Swiss Heart
Foundation. This work was supported by grant Nr. 205321-105597
from the Swiss National Science Foundation and by grant Nr.~8057.1
LSPP-LS from the Swiss Innovation Promotion Agency, CTI.
\end{acknowledgments}


\end{document}